\documentclass[conference]{IEEEtran}
\IEEEoverridecommandlockouts
\usepackage{cite}
\usepackage{amsmath,amssymb,amsfonts}
\usepackage{algorithmic}
\usepackage{graphicx}
\usepackage{textcomp}
\usepackage{xcolor}
\usepackage{multirow}
\usepackage{jabbrv}

\def\BibTeX{{\rm B\kern-.05em{\sc i\kern-.025em b}\kern-.08em
    T\kern-.1667em\lower.7ex\hbox{E}\kern-.125emX}}

\makeatletter
\newcommand{\linebreakand}{%
\end{@IEEEauthorhalign}
\hfill\mbox{}\par
\mbox{}\hfill\begin{@IEEEauthorhalign}
}
\makeatother

\begin{document}

\title{Enhanced Generative Adversarial Networks \\for Unseen Word Generation from EEG Signals
\footnote{{\thanks{This work was supported by Institute for Information \& Communications Technology Planning \& Evaluation (IITP) grant funded by the Korea government (MSIT) (No.2021-0-02068, Artificial Intelligence Innovation Hub; No. 2019-0-00079, Artificial Intelligence Graduate School Program(Korea University)).}
}}
}

\author{
\IEEEauthorblockN{Young-Eun Lee}
\IEEEauthorblockA{\textit{Dept.~of~Brain~and~Cognitive~Engineering}\\
\textit{Korea~University}\\
Seoul,~Republic~of~Korea\\
ye\_lee@korea.ac.kr}\\

\and

\IEEEauthorblockN{Seo-Hyun Lee}
\IEEEauthorblockA{\textit{Dept. of Brain and Cognitive Engineering} \\
	\textit{Korea University} \\
	Seoul, Republic of Korea \\
	seohyunlee@korea.ac.kr} \\

\and

\IEEEauthorblockN{Soowon Kim}
\IEEEauthorblockA{\textit{Dept. of Artificial Intelligence} \\
	\textit{Korea University} \\
	Seoul, Republic of Korea \\
	soowon\_kim@korea.ac.kr} \\

\linebreakand 

\IEEEauthorblockN{Jung-Sun Lee}
\IEEEauthorblockA{\textit{Dept. of Artificial Intelligence}\\
\textit{Korea University} \\
Seoul, Republic of Korea \\
jungsun\_lee@korea.ac.kr} \\

\and

\IEEEauthorblockN{Deok-Seon Kim}
\IEEEauthorblockA{\textit{Dept. of Artificial Intelligence} \\
\textit{Korea University} \\
Seoul, Republic of Korea \\
deokseon\_kim@korea.ac.kr} \\

\and

\IEEEauthorblockN{Seong-Whan Lee}
\IEEEauthorblockA{\textit{Dept. of Artificial Intelligence} \\
\textit{Korea University} \\
Seoul, Republic of Korea \\
sw.lee@korea.ac.kr}
}

\maketitle

\begin{abstract}
Recent advances in brain-computer interface (BCI) technology, particularly based on generative adversarial networks (GAN), have shown great promise for improving decoding performance for BCI. Within the realm of Brain-Computer Interfaces (BCI), GANs find application in addressing many areas. They serve as a valuable tool for data augmentation, which can solve the challenge of limited data availability, and synthesis, effectively expanding the dataset and creating novel data formats, thus enhancing the robustness and adaptability of BCI systems.
Research in speech-related paradigms has significantly expanded, with a critical impact on the advancement of assistive technologies and communication support for individuals with speech impairments.
In this study, GANs were investigated, particularly for the BCI field, and applied to generate text from EEG signals.
The GANs could generalize all subjects and decode unseen words, indicating its ability to capture underlying speech patterns consistent across different individuals. The method has practical applications in neural signal-based speech recognition systems and communication aids for individuals with speech difficulties.
\end{abstract}

\begin{small}
\end{small}

\section{INTRODUCTION}

In recent years, brain-computer interfaces (BCIs) have emerged as promising new areas of research, providing a means for humans to interact with external devices or control environments via brain signals~\cite{chaudhary2016brain}. Electroencephalography (EEG), which captures electrical activity by non-invasively attaching electrodes on the scalp, stands out among the several approaches for BCI research~\cite{thung2018conversion}. EEG offers an electrical measurement of cerebral activity without any surgical process of implanting electrodes, making it a significant source of information for applications involving brain signals~\cite{angrick2022towards}. BCIs based on EEG have been investigated for a variety of uses, including the control of motor activities, communication, and cognitive assessment~\cite{mane2021fbcnet, lee2019possible}. Despite the low signal-to-noise ratio of non-invasive EEG signals, several applications employing EEG have been explored owing to their ease of use and practical advantages~\cite{lee2020real}.

Generative adversarial networks (GANs) have emerged as a seminal framework in the field of machine learning, computer vision, and BCI. Originally proposed by Goodfellow et al.~\cite{goodfellow2014generative}, GANs have revolutionized the generation of realistic data by pitting a generator network against a discriminator network in a minimax game, resulting in the generation of data that closely mimics the underlying distribution. GANs have found wide-ranging applications, including image synthesis, speech synthesis, style transfer, and data augmentation, making them a fundamental tool in contemporary artificial intelligence research~\cite{bang2021spatio}.

Furthermore, GANs have found notable applications within the domain of EEG, where they are employed for both data augmentation and signal reconstruction~\cite{hartmann2018eeg}. In the context of EEG data, data augmentation techniques using GANs play a pivotal role in expanding the size and diversity of available datasets~\cite{panwar2020modeling, abdelfattah2018augmenting}. By generating synthetic EEG recordings that closely resemble real-world data, GANs enable improved training of machine learning models, thereby enhancing the models' ability to generalize and perform robustly on unseen data.
Additionally, GANs are leveraged in EEG signal reconstruction tasks, where the goal is to restore corrupted or missing segments of EEG recordings. This application is particularly valuable in scenarios where EEG data may be affected by noise or artifacts~\cite{lee2020reconstructing, sawangjai2021eeganet}, such as in clinical settings or during real-time monitoring~\cite{suk2014predicting}. GANs, through their generative capabilities, can effectively reconstruct EEG signals, contributing to improved signal quality and the extraction of meaningful information, which is crucial for various neuroscientific and clinical applications~\cite{lee2023AAAI, tang2023semantic}. The dual role of GANs in EEG data augmentation and signal reconstruction underscores their versatility and growing significance in the field of BCI.

\begin{figure*}[t]
\centering
    \includegraphics[]{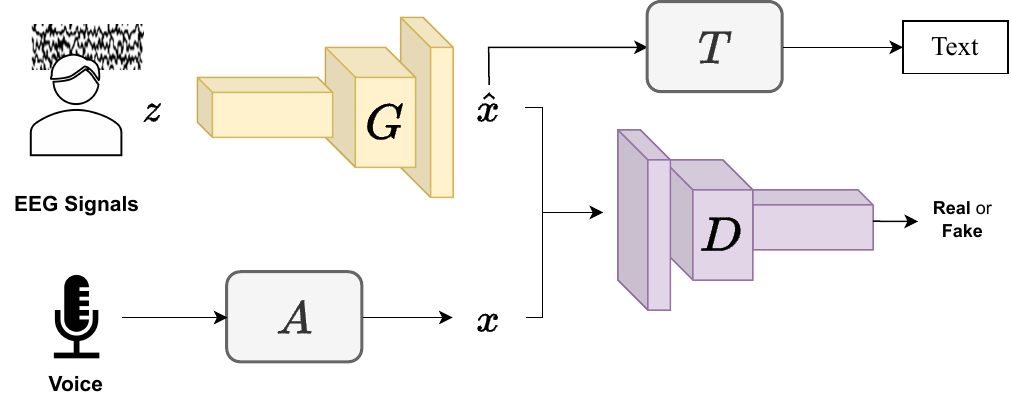}
    \caption{Brain-to-text system translates the brain signals into text. The character embedding is generated from the brain signals, given the target text generated from the voice. The overall framework of the GAN-based text generation model consists of a generator and discriminator with pretrained ASR. The generator converts the embedding vector $z$ of the EEG from multiple speakers into character embedding vector $x$. The discriminator distinguishes the character embedding vector as real or fake based on adversarial learning.}
    \label{fig1}
\end{figure*}

BCI-based communication systems have been developed in various ways, including exogenous paradigms such as event-related potential and steady-state visually evoked potential and intuitive paradigms such as spoken speech and imagined speech. 
Recent studies have investigated the use of brain signals including electrocorticography (ECoG) \cite{angrick2019speech}, and stereoEEG (sEEG) to reconstruct text or speech using machine learning and deep learning models. Willett et al. \cite{willett2021high} developed a novel handwriting communication system based on intracortical BCI by reconstructing pen trajectories of written alphabets using a recurrent neural network. ECoG studies have explored speech synthesis from brain signals, using linear and nonlinear regression approaches and recurrent neural networks during spoken/mimed speech \cite{herff2019generating}. Studies on sEEG have shown potential for reconstructing speech using a multiple input convolutional neural network or a unit selection approach \cite{anumanchipalli2019speech}. However, invasive techniques like as ECoG and sEEG entail risks and challenges to be applied for non-patient users. Therefore, non-invasive signals such as functional magnetic resonance imaging and EEG is gaining attention to be investigated, despite their inferior performance \cite{ross2022neural, lee2023AAAI, lee2019towards, kim2023diff}. Recently, Lee et al.~\cite{lee2023AAAI} demonstrated the potential of speech synthesis at the word level from EEG during spoken and imagined speech. This study has yielded promising findings; however, it is notable that the training time required for the proposed model was considerable, and a substantial amount of calibration data for each subject was required for optimal performance.

In this study, our primary objective was to explore the practical use of GANs within the field of BCI. Specifically, we focused on the intricate task of reconstructing text-based output directly from brain signals, aiming to advance the capabilities of BCI systems in generating textual information from neural data, thereby contributing to the broader domain of human-computer interaction and assistive technologies.

\section{MATERIALS AND METHODS}

\subsection{Generative Adversarial Networks}

In the context of our GAN-based approach, a generator network, denoted as $G$, learns to map input vectors $z$ from a latent space to synthetic data samples that resemble the target mel-spectrogram from voice $x$. The discriminator network, denoted as $D$, is simultaneously trained to distinguish between real brain signals and those generated by $G$. Through a min-max game, $G$ aims to generate increasingly realistic signals to deceive $D$, while $D$ strives to become more proficient at discrimination. This adversarial training process continues until an equilibrium is reached, where $G$ generates brain signals that are indistinguishable from real ones.

\subsection{Adversarial Loss}

The central component of the GAN framework is the adversarial loss, which comprises a generator loss ($L_G$) and a discriminator loss ($L_D$). The generator loss encourages $G$ to create synthetic signals that are convincingly realistic, while the discriminator loss guides $D$ in effectively distinguishing between real and synthetic data. The generator loss is minimized when $G$ produces signals that successfully fool the discriminator, and the discriminator loss is minimized when it accurately discriminates between real and generated signals. This competitive interplay between $L_G$ and $L_D$ drives the refinement of both networks, ultimately leading to the generation of brain signals that closely mimic the underlying distribution of real data.
Here is the equation of loss:
\begin{equation}
L_D = \mathbb{E}[\log(D(x))] + \mathbb{E}[\log(1 - D(G(z))],
\end{equation}
\begin{equation}
L_G = \mathbb{E}[\log(D(G(z))].
\end{equation}

\subsection{Dataset}

This study involved EEG signals and voice recordings of twenty-one subjects while they performed speech. The recordings were captured using scalp EEG consisting 64-channel electrode array placed on the scalp, in conjunction with a microphone to record voices, which was synchronized with the EEG signals. The subjects were instructed to speak in accordance with the instructions displayed on the screen. They spoke thirteen distinct words including ``ambulance," ``light," ``TV," ``water," ``pain," ``hello," ``toilet," ``clock," ``yes" ``stop," ``help me," ``thank you," and rest, in the same manner as the study \cite{lee2020neural}. Four trials were rhythmically given following each random auditory cue indicating the word or phrase the subject should speak. To minimize potential visual or auditory artifacts, the experiment was designed to capture EEG signals without any stimuli present. The study received approval from the Institutional Review Board and was conducted in accordance with the principles outlined in the Declaration of Helsinki. Prior to commencing the study, all subjects provided informed consent in adherence to ethical standards.

\subsection{Text Generation from Brain Signals}

Fig.~\ref{fig1} provides an overview of our proposed framework, which receives EEG signals from multiple speakers as input and generates corresponding text as output. It includes embedding processes, a generator, and a discriminator. 

For each subject, the dataset was split into five folds with a fixed random seed used for training, validation, and test sets. In order to make the unseen word for evaluating the scalability of the model, the word, `stop', was removed from the training dataset because all the phonemes composing the word were already included in the 11 words/phrases used for training. Thus, the model was trained using 11 words/phrases and a silent phase as the training set, while 12 words/phrases including the unseen word and a silent phase were used for validation and testing.

\subsection{Preprocessing}

Preprocessing methods were applied to the recorded raw EEG data, including downsampling to 250~Hz, band-pass filtering between 0.5 and 125~Hz, notch filtering at 60 and 120~Hz, and re-referencing using a common average reference method. To eliminate potential noise from movement and sound, automatic electrooculography and electromyography removal methods were employed to remove ocular and muscular artifacts \cite{kim2015abstract, gomez2006automatic}. Moreover, the EEG signals in the high-gamma frequency band were selected for training the model and data analysis. The dataset was epoched into 2-second segments and baseline correction was executed 500~ms before speaking.
Preprocessing procedures, as outlined in lee et al. \cite{lee2021mobile}, were implemented using MATLAB-based toolboxes including OpenBMI Toolbox \cite{leeMH2019eeg} and EEGLAB \cite{Delorme2004EEGLAB}.

The proposed framework employs embedding processes to generate representative feature vectors from EEG signals. This process involves using the common spatial pattern (CSP) technique and log-variance functions to extract spatio-temporal features from the EEG signals. CSP is a widely used mathematical approach in the area of BCI that separates a multivariate signal into subcomponents with the greatest variance fluctuations across two windows \cite{devlaminck2011multisubject}. Feature embedding was performed through time-wise computing of CSP patterns, with each EEG segment divided into 16 time points. The multi-CSP algorithm was used to calculate eight patterns per class, with the pattern size for each time point set to 104. The signals from each channel were processed to mean normalization to illustrate the temporal fluctuations of the embedding characteristics.

\begin{figure}[t]
\centering
    \includegraphics[]{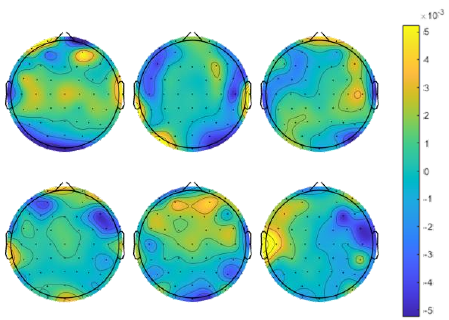}
    \caption{Spatial patterns of synchronization and de-synchronization of EEG signals for six different words. The color indicates the power of each channel across all trials.}
    \label{fig2}
\end{figure}

\section{RESULTS AND DISCUSSION}

\subsection{Results}

We evaluated the performance of our generated text from brain signals using the traditional metric: character error rate (CER). 
The performance of seen words and unseen words for generating text from brain signals was shown in CER. The CER of seen words was $61.8 \pm 8.5$ and the CER of unseen words was $83.3 \pm 3.9$.
We observed that our model was able to generalize to totally unseen subjects who were not included in the training set, demonstrating its ability to capture underlying patterns of speech that are consistent across different individuals. 
The results show that our method has the capacity to generalize to new subjects and speech patterns when given a small amount of calibration data, where data collection may be challenging due to limitations in sample size or data variability. 
Our findings demonstrate that the proposed model was able to learn global speech characteristics using data acquired from multiple speakers. Furthermore, our approach may have important practical applications, such as in the development of speech recognition systems or communication aids for individuals with speech impairments.

\subsection{Spatial Analysis}

To investigate differences in brain activity during speech, we conducted an analysis of spatial and spectral features, as shown in Fig.~\ref{fig2}. In the spatial characteristics, our analysis revealed that prominent synchronization in the central lobe and de-synchronization in the temporal lobe were observed \cite{Lee2019comparative}. In terms of the spectral features, we discovered that the high-gamma frequency range between 30~Hz to 120~Hz was dominantly synchronized, which is selected to analyze. In addition, we observed that the spatial characteristics of each text revealed significant differences, with each text exhibiting its distinct pattern of brain activation. These findings suggest that different aspects of speech may engage distinct neural networks, and the observed patterns of brain activation may reflect the cognitive and motor processes underlying speech production.

\subsection{Limitation and Future Works}

In our study, we employed a limited dataset consisting of only thirteen words, which offers a challenge in generating an extensive range of texts from brain signals. While we attempted to train the model to generate the unseen word by incorporating all the phonemes present in the unseen word, the outcomes were not satisfactory enough.
We plan to extend our research in the future by employing a sentence-based dataset that contains a large number of phonemes that should be capable of generating various words. In addition, we will utilize our proposed approach for generating text from the brain signals during the imagined speech, which may serve as a communication method for individuals who experience speech impairments~\cite{kim2019subject}.

\section{CONCLUSION}

In this paper, we investigated GAN for BCI, and presented the generative adversarial networks for BCI, in particular, for speech-related brain signals.
Our approach employs embedding processes, a generator, and a discriminator to learn the underlying patterns of speech and generate text at the word level with high fidelity. The model was able to learn global speech characteristics, demonstrating that our approach can generalize to unseen subjects and words, with a low error rate. 
Additionally, our spatial analysis revealed significant differences in brain activity during speaking, indicating distinct neural networks are engaged in different aspects of speech production. Despite the limited amount of words in the training dataset, our method has the potential to be applied in real-world applications such as speech recognition systems and communication aids for those with speech impairments.
Future research will focus on expanding our dataset to include more words and sentences, further improving the framework's generalization ability.

\bibliographystyle{jabbrv_IEEEtran}
\bibliography{REFERENCE}

\end{document}